\documentclass[Afour,times,sagev]{sagej}

\usepackage{moreverb,url}
\usepackage{mathabx}
\usepackage[ruled,linesnumbered]{algorithm2e}
\usepackage{bm}
\usepackage{multirow}
\usepackage{rotating}
\usepackage{float}
\usepackage{threeparttable}
\floatstyle{plaintop}
\usepackage{amsmath}
\usepackage{algpseudocode}
\floatname{algorithm}{New name for algorithm}
\makeatletter
\let\algorithmname\fname@algorithm
\makeatother

\usepackage[colorlinks,bookmarksopen,bookmarksnumbered,citecolor=blue,urlcolor=blue,linkcolor = blue]{hyperref}

\newcommand\BibTeX{{\rmfamily B\kern-.05em \textsc{i\kern-.025em b}\kern-.08em
T\kern-.1667em\lower.7ex\hbox{E}\kern-.125emX}}

\def\volumeyear{2023}

\begin{document}

\runninghead{Linke et al.}

\title{A Nonparametric Approach for Estimating the Effective Sample Size in Gaussian Approximation of Expected Value of Sample Information }

\author{Linke Li\affilnum{1,2}, Hawre Jalal\affilnum{3}, and Anna Heath\affilnum{1,2,4}}

\affiliation{\affilnum{1}Dalla Lana School of Public Health, University of Toronto, Toronto, Canada; \affilnum{2}Child Health Evaluative Sciences, The Hospital for Sick Children, Toronto, Canada; \affilnum{3}School of Epidemiology and Public Health, University of Ottawa, Ottawa, Canada; \affilnum{4}Department of Statistical Science, University College London, London, United Kingdom;}

\corrauth{Linke Li, Dalla Lana School of Public Health, University of Toronto, 155 College St Room 500, Toronto, Ontario, M5T 3M7, Canada.}

\email{link.li@mail.utoronto.ca}

\begin{abstract}

The effective sample size (ESS) measures the informational value of a probability distribution in terms of an equivalent number of study participants. The ESS plays a crucial role in estimating the Expected Value of Sample Information (EVSI) through the Gaussian approximation approach. Despite the significance of ESS, existing ESS estimation methods within the Gaussian approximation framework are either computationally expensive or potentially inaccurate. To address these limitations, we propose a novel approach that estimates the ESS using the summary statistics of generated datasets and nonparametric regression methods. The simulation results suggest that the proposed method provides accurate ESS estimates at a low computational cost, making it an efficient and practical way to quantify the information contained in the probability distribution of a parameter. Overall, determining the ESS can help analysts understand the uncertainty levels in complex prior distributions in the probability analyses of decision models and perform efficient EVSI calculations.

\end{abstract}


\maketitle

\section{INTRODUCTION}

The effective sample size (ESS) quantifies the informational value of a probability distribution by representing it in terms of an equivalent number of observations contributing information to this distribution.\cite{morita2008determining} For example, if the probability of successes $p$ follows a beta distribution with parameters (alpha = 2, beta = 8), the ESS is the number of successes (2) + the number of failures (8) = 10 observations. 
ESS can be particularly useful in computing the Expected Value of Sample Information (EVSI) using the Gaussian approximation (GA) approach.\cite{jalal2018gaussian} EVSI is a quantitative measure used in decision analysis to assess the potential value of acquiring additional data or information through a sample. It represents the expected improvement in decision-making or reduction in uncertainty that would result from collecting new data before making a choice or decision.  In addition to its importance in understanding the information in probability distributions and for EVSI estimation, ESS also finds various applications in clinical trial designs, including prior information elicitation, sensitivity analysis and trial protocol evaluations.  \cite{morita2008determining}

Despite its usefulness, existing ESS estimation methods within the GA have limitations.\cite{heath2020calculating} This is because these methods either entail potentially high computational costs, impeding their practicality, or exhibit inaccuracies, undermining the reliability of estimates.\cite{jalal2018gaussian} To efficiently assess the informative level of probability distributions and simplify the EVSI calculation, we propose a novel method for estimating ESS based on summary statistics and nonparametric regression models, inspired by the nonparametric regression-based EVSI methods proposed by Strong et al.\cite{strong2015estimating} We begin by reviewing the ESS, particularly within the GA approach, and proceed to illustrate the nonparametric regression-based ESS estimation method. Subsequently, we perform a simulation to demonstrate the effectiveness and accuracy of our method, followed by a brief discussion to conclude.

\section{METHODS}

\subsection{Effective Sample Size and Gaussian Approximation}

The effective sample size (ESS) is a statistical concept that quantifies the informational value of a probability distribution through the number of hypothetical participants \cite{morita2008determining}. A greater ESS signifies a more informative distribution, while a lower value indicates the opposite. The estimation of ESS serves two important purposes in health economic evaluation. Firstly, it allows area experts to determine if the amount of information contained in the distribution aligns with their prior beliefs. Secondly, it facilitates the estimation of the EVSI within the GA approach.\cite{jalal2018gaussian, linke2023estimating}
This efficiency stems from the characteristic of ESS as being intrinsic to the GA setup, where it is computed only once. After that, this value remains unaltered for EVSI calculations across a spectrum of study designs, irrespective of sample size variations. \footnote{To estimate EVSI, ESS is combined with the nonparametric regression models and Taylor series expansions to efficiently approximate the distribution of the conditional economic benefit and estimate EVSI. We refer readers to the article by Li et al. for a comprehensive explanation of the EVSI estimation method.\cite{linke2023estimating}} This section introduces the concept of ESS as it is presented within the GA approach.  

Firstly, assume we have a parameter $\phi$ with a corresponding probability distribution $p(\phi)$ that can be well approximated by a Gaussian distribution. This distribution has mean $\mu_0$ and variance $\frac{\sigma^2}{n_0}$:

\begin{equation}
\phi \sim N\left(\mu_0,\frac{\sigma^2}{n_0}\right), \label{eq:prior}
\end{equation}
where $\sigma^2$ represents the individual-level variance of the data collection process, and $n_0$ can be understood as the number of hypothetical participants that would yield the uncertainty level of $\phi$. $n_0$ is defined as the ESS of $p(\phi)$ in the GA approach.

Additionally, we denote the set of $n$ observations obtained from the data collection exercise that would provide additional information on $\phi$ as $\bm{X}_n$, and its sample mean as $\widebar{\bm{X}}_n$. In the GA approach, it is assumed that given the parameter $\phi$,  $\widebar{\bm{X}}_n$ is approximately Gaussian distributed  with mean $\phi$ and variance $\frac{\sigma^2}{n}$:
\begin{equation}
\widebar{\bm{X}}_n |\phi \sim N\left(\phi,\frac{\sigma^2}{n}\right). \label{eq:likelihood}
\end{equation}

Due to the conjugate nature of the distributions of $\phi$ and $\widebar{\bm{X}}_n |\phi$, the conditional mean of $\phi$ given the observed data is a weighted combination of   $\mu_0$ and $\widebar{\bm{X}}_n$: \begin{equation}
\mathbb{E}_{\phi}[\phi|\bm{X}_n] = (1 - v) \mu_0 + v \widebar{\bm{X}}_n, v =  \frac{n }{ n_0 + n}. \label{eq:posteior_mean_eq}
\end{equation}

Using this formulation, the ESS $n_0$ can be determined based on the variance of either $\mathbb{E}_{\phi}[\phi|\bm{X}_n]$ or $\widebar{\bm{X}}_n$. This is because the marginal distributions of $\mathbb{E}_{\phi}[\phi|\bm{X}_n]$ and $\widebar{\bm{X}}_n$ are both Gaussian with variances $\frac{v \sigma^2}{n_0}$ and $\frac{\sigma^2}{v n_0}$, respectively. \cite{jalal2018gaussian} As the variance of $p(\phi)$ is $\frac{\sigma^2}{n_0}$, 
we can estimate $n_0$ using variance ratios. Specifically, the variance ratio between $\phi$ and $\mathbb{E}_{\phi}[\phi|\bm{X}_n]$ yeilds an $n_0$ estimation of:

\begin{equation}
\hat n_0 = n\left ( \frac{\mbox{Var}_{\phi}[\phi]}{\mbox{Var}_{\bm{X}_n}\left[\mathbb{E}_{\phi}[\phi|\bm{X}_n] \right]} - 1 \right ), \label{eq:n_0_formula_2} 
\end{equation}
and the variance ratio between $\phi$ and $\widebar{\bm{X}}_n$ yeilds an $n_0$ estimation of:

\begin{equation}
\hat n_0 = n\left ( \frac{\mbox{Var}_{\widebar{\bm{X}}_n}[\widebar{\bm{X}}_n]}{\mbox{Var}_{\phi}[\phi]} - 1 \right ). \label{eq:n_0_formula}
\end{equation}
Although equations  \ref{eq:n_0_formula_2}  and \ref{eq:n_0_formula} are derived using the Gaussian assumption, Jalal and Alarid-Escudero have shown that they can be generalized to estimate $n_0$ for other non-Gaussian likelihood and prior combinations.  \cite{jalal2018gaussian}  


\subsection{Estimate the Effective Sample Size using the Nonparametric Regression-based Method }\label{ch:compute_n_0}

In this section, we review current $n_0$ estimation methods and note their drawbacks.\cite{jalal2018gaussian}  We then introduce a novel approach, which employs nonparametric regression models, to efficiently and accurately estimate $n_0$.

In the original GA article, Jalal and Alarid-Escudero introduced two methods that can be generalized to estimate $n_0$ for nonconjugate priors. \cite{jalal2018gaussian}  The first method uses equation \ref{eq:n_0_formula_2}. For this, we need to estimate $\mathbb{E}_{\phi}[\phi|\bm{X}_n]$ for a large number of datasets $\bm{X}_n$ generated from a specific data collection exercise. For each dataset, we can use Markov Chain Monte Methods (MCMC) to simulate from the posterior distribution of parameter $\bm\phi$ and calculate the mean. We then compute the sample variance of these estimated posterior means to approximate $\mbox{Var}_{\bm{X}_n}\left[\mathbb{E}_{\phi}[\phi|\bm{X}_n] \right]$ and estimate $n_0$.
While this method yields accurate $n_0$ estimates, repeated MCMC evaluations may impose a significant computational burden and complicate practical implementation.

The second method uses equation \ref{eq:n_0_formula}. However, since the sample mean $\widebar{\bm{X}}_n$ may not have the same scale as $\phi$ for many non-Gaussian probability distributions, equation \ref{eq:n_0_formula} can not be directly applied. In such cases, we need to identify other summary statistics $S$ that share the same scale as $\phi$ and compare its variance with that of $\phi$ to estimate  $n_0$: 
\begin{equation}
\hat n_0 = n\left ( \frac{\mbox{Var}_{S}[S]}{\mbox{Var}_{\phi}[\phi]} - 1 \right ). \label{eq:n_0_formula_3} 
\end{equation}
As this method only requires summary statistics, it is more computationally efficient than MCMC. However, it necessitates the identification of suitable summary statistics for $\bm{X}_n$. These statistics should exhibit the same scale as $\phi$ and have an appropriate level of variation, which can pose challenges in real-world case studies.

To reduce the computational burden of $n_0$ when the appropriate summary statistics of $\phi$ are difficult to find, we suggest estimating $\mathbb{E}_{\phi}[\phi|\bm{X}_n]$ using the nonparametric regression-based method, motivated by the EVSI method proposed by Strong et al. \cite{strong2015estimating,strong2014estimating} To achieve this, we need to first determine a low-dimensional summary statistic of the dataset $\bm{X}_n$, denoted $T(\bm{X}_n)$. Note that, different from the current summary statistics-based method, the regression-based method does not require   $T(\bm{X}_n)$ to be converted to the same scale as $\phi$, which makes it simpler to specify.  A commonly used summary statistic for the nonparametric regression-based method is the maximum likelihood estimate (MLE) of $\phi$ given $X_{n}$.

After we find $T(\bm{X}_n)$, using the formulation suggested by Strong et al., $\phi$ can be expressed as the function of $T(\bm{X}_n)$, plus a zero mean error term $\epsilon$ \cite{strong2015estimating}
\begin{equation}
\phi =  g(  T(\bm{X}_{n})) + \epsilon. \label{eq:nonpar_1}
\end{equation} 
This equation suggests that the conditional expectation $\mathbb{E}_{\phi}[\phi|\bm{X}_n]$ can be approximated by the fitted value of a nonparametric regression model which takes $\phi$ as the response and $ T(\bm{X}_{n})$ as the predictor. Therefore, our proposed method begins by fitting this non-parametric regression and then extracting the fitted values from this regression model. Next,  $\mbox{Var}_{\bm{X}_n}\left[\mathbb{E}_{\phi}[\phi|\bm{X}_n] \right]$ is approximated by the variance of these fitted values. Finally, $n_0$ can then be estimated from equation  \ref{eq:n_0_formula}. The detailed algorithm for estimating $n_0$ using the nonparametric regression-based method is summarized in box \ref{alg:ESS_Taylor}.
 
\begin{algorithm}

\caption{Nonparametric Regression-Based Effective Sample Size Estimation Algorithm}\label{alg:ESS_Taylor}

\begin{algorithmic}[1]
\State \textbf{Generate samples of 
parameters and datasets}:
\begin{itemize}
    \item Generate $M$ samples of parameter $\phi$ from the probabilistic distribution $p(\phi)$, denoted $\phi^1, \dots, \phi^M$. 
    \item For $m$ from $1$ to $M$, use the sample $\phi^m$ to generate a sample of the dataset at the sample size $n$, $\bm{X}_n^m$, and compute its summary statistics $T(\bm{X}_n^m)$.
\end{itemize}
\State \textbf{Fit the regression model and estimate  the effective sample size}:
\begin{itemize}
    \item Regress the parameter samples $\phi^m$ on the summary statistics $T(\bm{X}_n^m)$ using a nonparametric regression model.
    \item Extract the fitted values from the regression model and calculate their sample variance, denoted  $\widehat{\mbox{Var}}_{\bm{X}_n}\left[\mathbb{E}_{\phi}[\phi|\bm{X}_n] \right]$.
    \item Calculate the sample variance of $\phi^1, \dots, \phi^M$, denoted $\widehat{\mbox{Var}}_{\phi }\left[\phi\right]$.
    \item Estimate the effective sample size of $\phi$ using $n\left ( \frac{\widehat{\mbox{Var}}_{\phi}[\phi]}{\widehat{\mbox{Var}}_{\bm{X}_n}\left[\mathbb{E}_{\phi}[\phi|\bm{X}_n] \right]} - 1 \right )$.

\end{itemize}

\end{algorithmic}
\end{algorithm}

\subsection{Case study: Examing the Accuracy of the Nonparametric Regression-based $n_0$ Estimation }

We assess the accuracy of the proposed method across seven diverse data collection scenarios, including settings where $\phi$ is univariate and multivariate.  The distributions chosen for prior and data generation, respectively, for seven scenarios are: 1) beta-binomial, 2) gamma-exponential, 3) Poisson-gamma, 4) Gaussian-Weibull, 5) Dirichlet-multinomial, 6) truncated normal-binomial and 7) transformed beta-exponential. For each scenario, we compute $n_0$ using the original summary statistic and MCMC methods, and our nonparametric regression-based method. Moreover, for scenarios one to four, where closed-form posterior distributions exist, we derive the analytic value of $n_0$. Table \ref{tab:case_1} details the prior distributions, likelihood functions, and $n_0$ for each scenario.

To calculate the ESS using our nonparametric regression-based method, we generate $10^4$ parameter samples from each prior distribution. Each parameter sample is then inputted into the likelihood function of the study design to generate $10^4$ datasets. Subsequently, we compute the maximum likelihood estimate (MLE) as the summary statistic for each dataset. For our method, we then regressed the parameter samples on the MLEs using a spline. The fitted values were extracted from the regression model and used to compute $n_0$ according to equation \ref{eq:n_0_formula}. The summary statistics-based method also used the MLEs to estimate $n_0$ by equation \ref{eq:n_0_formula_3}, while the MCMC method was based on $10^4$ posterior samples for each dataset to estimate $\mathbb{E}_{\phi}[\phi|\bm{X}_n]$.\cite{jalal2018gaussian}

\section{Results}

Table \ref{tab:case_1} provides the estimated $n_0$ values and the associated time to compute the ESS for each data collection process using distinct estimation methods. 
For the data collection exercises one to four, the estimates of $n_0$ obtained from the nonparametric regression-based methods consistently align with the analytic results. In contrast,  the summary statistics-based method overestimates the $n_0$ for the second experiment. Although the analytic result for $n_0$ is unavailable for the data collection exercises five to seven, the estimates provided by the nonparametric regression-based method remains consistent with that obtained from the MCMC method. Conversely, the summary statistics-based method may overestimate $n_0$ for these experiments. 
In terms of computational efficiency, both the nonparametric regression-based method and the summary statistics-based method can estimate $n_0$ within seconds or minutes, whereas the MCMC method takes about an hour in similar settings.

\begin{table*}[t]
\centering

\caption{ The probability distribution, likelihood function and effective sample size ($n_0$) derived using the analytic results for five data collection exercises. The $n_0$ estimates and associated computational time given by the summary statistics approach, Nonparametric regression-based approach and MCMC approach are also reported.\label{tab:case_1}}

\begin{tabular}{cccc}
\toprule
 \multicolumn{4}{l}{\qquad  \textbf{Simulation Settings }}  \\\cmidrule{1-4}
 &\textbf{Experiment 1} & \textbf{Experiment 2} &\textbf{Experiment 3}  \\
\midrule
\text{Prior }&  \begin{tabular}{@{}c@{}}$P \sim Beta(\alpha = 4,$ \\$\beta = 6)$  \end{tabular}   &  \begin{tabular}{@{}c@{}}$\lambda \sim Gamma(\alpha  = 20,$ \\$ \beta = 10)$  \end{tabular}&\begin{tabular}{@{}c@{}}$\lambda \sim Gamma(\alpha  = 50,$ \\$\beta = 100)$  \end{tabular} \\
\text{Likelihood}&\begin{tabular}{@{}c@{}}$X \sim Binomial(n = 20, P)$ \end{tabular}  &\begin{tabular}{@{}c@{}}$X_i \sim Exp(\lambda)$ \\ $ i = 1, \dots, 100$ \end{tabular}  &   \begin{tabular}{@{}c@{}}$X_i \sim Possion(\lambda)$ \\ $ i = 1, \dots, 100$ \end{tabular} \\
\text{ Analytic $n_0$ }  &$n_0 = \alpha + \beta = 10$ & $n_0 = \alpha = 20$ & $n_0 = \beta = 100$ \\

\midrule
 \multicolumn{4}{l}{\qquad \quad \,  $\mathbf{n_0}$ \textbf{Estimates }} 
 
\\\cmidrule{1-4}
\text{Summary Statistics Method }  &$10.09$ & $24.17$ &$100.15$\\
\text{Regression-based Method}  &$9.93$& $21.14 $ & $ 99.58$ \\
\text{MCMC Method}  &$9.66$& $23.38 $ & $98.91$\\

\midrule
 \multicolumn{4}{l}{ \textbf{Computational Time of} $\mathbf{n_0}$ \textbf{(min)} }  
 
\\\cmidrule{1-4}

\text{Summary Statistics method }  &$0.01$& $0.01$ & $0.01$  \\
\text{Regression-based method}  &$0.01$& $0.01$ & $0.01$\\
\text{MCMC method}  &$78.41$& $49.98$ & $101.93$ \\

\bottomrule

\end{tabular}

\end{table*}

\begin{table*}[t]
\centering

\begin{tabular}{ccc}
\toprule
 \multicolumn{3}{l}{\qquad  \textbf{Simulation Settings }}  \\\cmidrule{1-3}
 &\textbf{Experiment 4}  &\textbf{Experiment 5} \\
\midrule
\text{Prior }&  \begin{tabular}{@{}c@{}} $P \sim Dirichlet(\bm{\alpha} = $ \\$ (10, 5, 8) )$  \end{tabular}   & \begin{tabular}{@{}c@{}}$\theta \sim N(\mu_0  = 1,$ \\$\sigma^2 = 0.04)$  \end{tabular}  \\
\text{Likelihood}& \begin{tabular}{@{}c@{}}$X \sim Multinomial(n = 50, P )$ \\  \end{tabular}  & \begin{tabular}{@{}c@{}}$X_i \sim Weibull(v  = 1,\lambda = 1/\theta)$  \\ $ i = 1, \dots, 100$ \end{tabular} \\
\text{ Analytic $n_0$ }  & $n_0 = 10+ 5+ 8 = 23 $ & --- \\

\midrule
 \multicolumn{3}{l}{\qquad \quad \,  $\mathbf{n_0}$ \textbf{Estimates }}  
 
\\\cmidrule{1-3}
\text{Summary Statistics Method }  &$23.26$  & $30.57$ \\
\text{Regression-based Method}   & $22.74$& $24.94$ \\
\text{MCMC Method}  &$22.67$   &$26.34$\\

\midrule
 \multicolumn{3}{l}{ \textbf{Computational Time of} $\mathbf{n_0}$ \textbf{(min)} }  
 
\\\cmidrule{1-3}

\text{Summary Statistics method }   &$0.01$  &  $3.45$\\
\text{Regression-based method}   & $0.01$  &$3.52$\\
\text{MCMC method}   & $150.55$ & $92.63$\\

\bottomrule

\end{tabular}

\end{table*}

\begin{table*}[t]
\centering

\begin{tabular}{ccc}
\toprule
 \multicolumn{3}{l}{\qquad  \textbf{Simulation Settings }}  \\\cmidrule{1-3}
 &\textbf{Experiment 6}  &\textbf{Experiment 7} \\
\midrule
\text{Prior }&  \begin{tabular}{@{}c@{}} $P \sim  TN(\mu_0  = 0.2,\sigma^2 = 0.01,$ \\$ a = 0, b = 1)$ \footnotemark  \end{tabular}    &  \begin{tabular}{@{}c@{}}$\lambda = -\log(1-P),$\\$P \sim Beta(\alpha = 4,\beta = 6)$ \footnotemark  \end{tabular}  \\
\text{Likelihood}& \begin{tabular}{@{}c@{}}$X \sim Binomial(n = 20, P)$ \\  \end{tabular}  & \begin{tabular}{@{}c@{}}$X_i \sim Exp(\lambda)$ \\ $ i = 1, \dots, 100$ \end{tabular}  \\
\text{ Analytic $n_0$ }  & --- & --- \\

\midrule
 \multicolumn{3}{l}{\qquad \quad \,  $\mathbf{n_0}$ \textbf{Estimates }}  
 
\\\cmidrule{1-3}
\text{Summary Statistics Method }  &$18.26$  & $7.40$ \\
\text{Regression-based Method}   & $16.67$& $4.87$ \\
\text{MCMC Method}  &$16.19$   &$4.90$\\

\midrule
 \multicolumn{3}{l}{ \textbf{Computational Time of} $\mathbf{n_0}$ \textbf{(min)} }  
 
\\\cmidrule{1-3}

\text{Summary Statistics method }   &$0.01$  &  $0.01$\\
\text{Regression-based method}   & $0.01$  &$0.01$\\
\text{MCMC method}   & $13.66$ & $1105.4$\\

\bottomrule

\end{tabular}

\begin{tablenotes}
\item $\dagger$ The prior distribution for the parameter $P$ follows a truncated normal distribution with a domain ranging from $0$ to $1$, a mean of $0.2$, and a variance of $0.01$.
\item $\ddagger$ The prior distribution of the parameter $\theta$ is defined as the negative natural logarithm of  $1 - P$, where  $P$ follows a beta distribution with parameters $\alpha = 4$ and $\beta = 6$.

\end{tablenotes}

\end{table*}
\addtocounter{footnote}{-2}

\section{Discussion}
This article introduces a novel computational method for efficiently quantifying the amount of information contained in a probability distribution. By combining summary statistics and nonparametric regression models, our method provides an estimation of the effective sample size (ESS) for various probability distributions based on a Gaussian approximation. 
The accuracy and efficiency of the proposed method are validated through a comprehensive simulation study, encompassing various types of probability distributions.

In addition to EVSI calculation, our efficient ESS estimation approach can also be used in other applications. This encompasses evidence synthesis in clinical trials, as explored by Morita et al.\cite{morita2008determining}, and informing parameter distributions in medical decision-making. For instance, 
our method can be particularly useful when the parameter's distribution is complex and ESS cannot be readily determined from the input's parameters.  In such instances,  ESS provides a valuable metric to quantify the number of participants contributing information to a specific input. This insight can be compared with an expert's intuition, and if there are discrepancies, the uncertainty representation can be revised accordingly.

Overall, ESS is a useful metric that can be readily used to inform EVSI and guide input distributions in medical decision-making. Our proposed approach can help estimate ESS more accurately and efficiently.

\begin{acks}
Placeholder.
\end{acks}

\begin{sm}
Supplementary material for this article is available on the Medical Decision Making Web site at http://journals.sagepub.com/home.mdm
\end{sm}

\bibliographystyle{SageV}
\bibliography{MDM_ref.bib}

\begin{thebibliography}{1}
\providecommand{\url}[1]{\texttt{#1}}
\providecommand{\urlprefix}{URL }
\expandafter\ifx\csname urlstyle\endcsname\relax
  \providecommand{\doi}[1]{DOI:\discretionary{}{}{}#1}\else
  \providecommand{\doi}{DOI:\discretionary{}{}{}\begingroup \urlstyle{rm}\Url}\fi
\providecommand{\eprint}[2][]{\url{#2}}

\bibitem{morita2008determining}
Morita S, Thall PF and M{\"u}ller P.
\newblock Determining the effective sample size of a parametric prior.
\newblock \emph{Biometrics} 2008; 64(2): 595--602.

\bibitem{jalal2018gaussian}
Jalal H and Alarid-Escudero F.
\newblock A gaussian approximation approach for value of information analysis.
\newblock \emph{Medical Decision Making} 2018; 38(2): 174--188.

\bibitem{heath2020calculating}
Heath A, Kunst N, Jackson C et~al.
\newblock Calculating the expected value of sample information in practice: considerations from 3 case studies.
\newblock \emph{Medical Decision Making} 2020; 40(3): 314--326.

\bibitem{strong2015estimating}
Strong M, Oakley JE, Brennan A et~al.
\newblock Estimating the expected value of sample information using the probabilistic sensitivity analysis sample: a fast, nonparametric regression-based method.
\newblock \emph{Medical Decision Making} 2015; 35(5): 570--583.

\bibitem{linke2023estimating}
Linke L, Hawre J and Anna H.
\newblock Estimating the expected value of sample information across different sample sizes using gaussian approximations and spline-based taylor series expansions.
\newblock \emph{Medical Decision Making} 2023; XX(X): XXX--XXX.

\bibitem{strong2014estimating}
Strong M, Oakley JE and Brennan A.
\newblock Estimating multiparameter partial expected value of perfect information from a probabilistic sensitivity analysis sample: a nonparametric regression approach.
\newblock \emph{Medical Decision Making} 2014; 34(3): 311--326.

\end{thebibliography}

\end{document}